\documentclass[twocolumn,aps,prl,epsf]{revtex4}

\usepackage{amsmath} 
\usepackage{amsfonts} 
\usepackage{epsfig} 
\usepackage{epsf} 
\usepackage{array}

\begin{document}

\title[Surface magnetic phase transition of the double-exchange ferromagnet]
{Surface magnetic phase transition of the double-exchange ferromagnet: Schwinger-boson mean-field study} 
\author{Satoshi Okamoto}
\altaffiliation[Electronic address: ]{okapon@ornl.gov} 
\address{Materials Science and Technology Division, Oak Ridge National Laboratory, Oak Ridge, Tennessee 37831, USA}

\begin{abstract}
The surface magnetic phase transition of a double-exchange model for metallic manganites 
is studied using a Schwinger-boson mean-field method. 
About three unit-cells wide surface layers are identified. 
The magnetic moment in these layers decreases more rapidly than that in the bulk when the temperature is increased. 
This behavior is consistent with experimental observations. 
We also discuss the implication of this behavior on the tunneling magnetoresistance effect using manganites and 
possible improvement of the magnetoresistance effect near the bulk Curie temperature. 
\end{abstract}

\maketitle


\section{Introduction}

Heterostructures involving transition-metal oxides have been attracting interest \cite{Izumi01,Ohtomo02,Okamoto04,Chkhalian06,Chakhalian07} 
because they provide a playground to explore new and useful functionalities that are not realized in the bulk. 
In addition, these are supposed to become fundamental building blocks of electronic devices utilizing a variety of properties of transition-metal oxides \cite{Imada98}. 
Among these oxides, perovskite manganites are promising candidates for spintronic devices due to their high spin polarization, 
high ferromagnetic Curie temperature ($T_C$), and the colossal magnetoresistance effect.

One potential application of perovskite manganites is as a tunneling magnetoresistance (TMR) junction \cite{Julliere75,Maekawa82}. 
The TMR junction consists of two ferromagnetic metallic leads separated by an insulating barrier. 
The conductance across the barrier can be changed by changing the relative orientation of magnetic moments of the two leads. 
Since high polarization in the ferromagnetic metallic leads is required to obtain the large magnetoresistace ratio (MR), 
highly-polarized perovskite manganites could serve as ideal ferromagnetic leads. 
Furthermore, electronic devices typically function at room temperature, therefore high $T_C$ materials  
are particularly favorable, such as La$_{1-x}$Sr$_x$MnO$_3$ with $0.2 < x < 0.5$ where $T_C$ reaches 350~K\cite{Urushibara95}.

Several attempts have been made to fabricate perovskite TMR junctions \cite{Viret97,Jo00,Bowen03}. 
A very large MR was measured at low temperature, consistent with half-metallicity. 
However, the MR decreases rapidly and disappears well below $T_C$ \cite{Bowen03}. 
On the basis of spin-resolved photoemission spectroscopy, 
it was suggested that the rapid decrease of MR is due to the stronger temperature dependence of the spin polarization at interfaces 
than in the bulk \cite{Park98}.

Surface magnetism has been theoretically studied within a classical Heisenberg model using the numerical Monte Carlo (MC) technique. 
The surface polarization was shown to decrease more rapidly than that in the bulk \cite{Binder74,Calderon99}. 
More recently, interfacial phase transition of the double-exchange (DE) model for manganites 
was studied by the dynamical-mean-field method \cite{Lin06} and the MC method \cite{Yunoki08}. 
Since the dynamical-mean-field theory (DMFT) neglects spatial correlations \cite{Georges96}, 
it is expected to become less accurate in low dimensional systems, and therefore at surfaces and interfaces. 
The MC requires a very large system to investigate surface or interface phase transitions. 
In fact, one-dimensional systems were used in \cite{Yunoki08}. 
Therefore, the difference between the bulk magnetism and the interface magnetism remains unresolved.

In this paper, we investigate the surface magnetic phase transition of the DE model by using an alternative technique, 
the Schwinger-boson mean-field (SBMF) method. 
We focus on metallic manganites possessing a relatively high $T_C$ such as La$_{1-x}$Sr$_x$MnO$_3$ with a doping concentration of $x \sim 0.3$. 
The SBMF method has had success in describing the behavior of the quantum Heisenberg model in low dimensions \cite{Arovas88} 
satisfying the Mermin-Wagner theorem \cite{Mermin66}. 
The SBMF method was also applied to the bulk DE model \cite{Sarker96,Arovas98}. 
Since the SBMF method correctly describes low-dimensional spin systems, 
it is also expected to provide a suitable description of surface and interface magnetic behavior. 

The paper is organized as follows. In section 2, the theoretical model and the formalism are outlined. 
In section 3, we present the numerical results, and in section 4, we discuss the implication on the TMR effect and summarize.

\section{Model and formalism}

Let us start by setting up our theoretical model. 
Perovskite manganites are known to exhibit a variety phenomena including the colossal magnetoresistance effect, 
charge/orbital orderings and nanoscale inhomogeneity. 
In order to understand all these phenomena, many effects must be considered. 
The kinetics of $e_g$ electrons $H_K$ and the Hund coupling (between the $e_g$ electrons and $t_{2g}$ spins) $H_H$ 
are the key ingredients in generating the DE interaction, resulting in metallic ferromagnetic states. 
In addition, there are 
electron-electron interactions $H_{e-e}$, orbital degrees of freedom, 
electron-lattice interactions $H_{e-l}$ (Jahn-Teller coupling is also included here), 
and chemical inhomogeneity. 
Including all the effects is certainly necessary to investigate the complicated multiphase behavior of manganites. 
On the other hand, our main interest is the surface magnetic behavior of high $T_C$ manganites. 
Although the quantitative agreement between theory and experiment remains incomplete, 
the characteristic behavior of high $T_C$ manganites can be understood based on the simple DE model \cite{Furukawa98,Millis96}. 
Furthermore at this moment, detailed information on the surface structure is not available, 
although, the recent work suggests that the surface layer of La$_{1-x}$Sr$_x$MnO$_3$ is a (La,Sr)O plane \cite{Bertacco02}. 
Therefore, we focus on the effect which highlights the difference between the bulk and the surface most: 
smaller coordination at the surface, 
by terminating the perfect cubic lattice at the [001] plane. 
For simplicity, we consider a single-orbital DE model, which reproduces many of the properties of metallic La$_{1-x}$Sr$_x$MnO$_3$ \cite{Furukawa98}, 
with infinite Hund coupling. 
We will also present brief discussion on the orbital polarization and disorder at the surface in terms of the transfer anisotropy. 
Since the theoretical model is rather simple, our discussion will be done on the qualitative level.

The Hamiltonian for the DE model is written as 
$
H = - t \sum_{\langle i j \rangle \sigma} \bigl( s_{i \sigma}^\dag s_{j \sigma} f_i^\dag f_j + H.c. \bigr).
$
Here, $t$ is the nearest-neighbor transfer, 
$s_{i \sigma}$ the spinor boson (Schwinger boson) with the local constraint $\sum_\sigma s_{i\sigma}^\dag s_{i\sigma} =1$ 
representing the rotation of spin space, and 
$f_i$ the spinless fermion representing an electron whose spin is parallel to the local moment 
$\vec S_i = S \sum_{\alpha \beta} s_{i \alpha}^\dag \vec \sigma_{\alpha \beta} s_{i \beta}$.  
%
The Lagrangian for this system becomes 
\begin{eqnarray}
L &=&
\sum_i \overline f_i (\partial_\tau - \mu) f_i + \sum_{i \sigma} \bigl( \overline f_i f_i + 2S \bigr)s_{i \sigma}^* \partial_\tau s_{i \sigma} \nonumber \\
&&
- t \! \sum_{\langle i j \rangle \sigma} \bigl( s_{i \sigma}^* s_{j \sigma} \overline f_i f_j + H.c. \bigr) \nonumber \\
&& 
+ \sum_i \lambda_i (\tau) \biggr(\sum_\sigma |s_{i \sigma}|^2 - 1 \biggl), 
\end{eqnarray}
where $\tau$ is the imaginary time, 
$\mu$ is the chemical potential for spinless fermions, 
$2S$ originates from the Berry phase of a localized spin $\vec S_i$, 
and the last term represents the local constraint $\sum_\sigma |s_{i\sigma}|^2=1$ 
with the Lagrange multiplier $\lambda_i (\tau)$.

At this stage, we introduce a mean-field approximation: 
$n_i = \langle \overline f_i f_{i} \rangle$, $\chi^s_{ij} = \langle \overline f_i f_j \rangle$, and 
$\chi^f_{ij} = \sum_\sigma \langle s_{i \sigma}^* s_{j \sigma} \rangle$, 
and relax the local constraint to the global one by neglecting the $\tau$ dependence of $\lambda$. 
By rescaling the spinor boson $s_{i\sigma}$ as $\sqrt{2S+n_i} \, s_{i \sigma} \Rightarrow s_{i \sigma}$, 
we obtain the mean-field Lagrangians for fermions and bosons 
\begin{equation}
L_f =
\sum_i \overline f_i (\partial_\tau - \mu) f_i 
- \frac{t}{2} \sum_{\langle i j \rangle} \left( \frac{\chi^f_{ij} \overline f_i f_j}{\sqrt{S_i^{tot} S_j^{tot}}}  + H.c. \right), 
\end{equation}
and
\begin{eqnarray}
L_s &=& 
\sum_{i \sigma} s_{i \sigma}^* \partial_\tau s_{i \sigma} - \frac{t}{2} \sum_{\langle i j \rangle \sigma} 
\left( \frac{\chi^s_{ij} s_{i \sigma}^* s_{j \sigma}}{\sqrt{S_i^{tot} S_j^{tot}}} + H.c. \right) \nonumber \\
&&+ \sum_i \lambda_i \biggr( \sum_\sigma |s_{i \sigma}|^2 - 2 S_i^{tot} \biggl), 
\end{eqnarray}
respectively, 
with $S_i^{tot} = S + n_i/2$. 
The order parameter $\chi^f$ is defined in the same way as before but now using the rescaled bosons. 
The local density of spinor bosons is now $2S_i^{tot}$. 
In this SBMF, $\chi^f_{ij}$ represents the nearest-neighbor ferromagnetic correlation, and 
the ordered moment $M$ is represented by the Bose condensation of spinors $N_0$ as $M=N_0/2$.

We solve the self-consistency equations numerically for an $N$ layer system with 
the open-boundary condition in the $z$ direction and the periodic-boundary condition in the $xy$ plane. 
Thus, $\chi^{f}$ and $\chi^{s}$ are dependent on the layer coordinate $z$ and the interplane bond, 
and $\lambda$ on $z$. 
In this mean-field theory, an additional phase transition appears above $T_C$ associated with 
the order parameters $\chi^{f,s}$. 
Since this phase transition is an artifact of the present decoupling scheme, we focus on the temperature range below $T_C$. 
The most difficult part lies in fixing $\{\lambda_i\}$ 
so that the constraint $\sum_\sigma |s_{i\sigma}|^2=2S_i^{tot}$ is satisfied at each layer. 
Note, $\sum_\sigma |s_{i\sigma}|^2$ at each layer depends on all $\lambda$s, and the simple bisection method does not work. 
Here, we apply the conjugate gradient algorithm and minimize the function 
$\Delta (\{ \lambda_i \}) = \sum_{l=1}^N \bigl|\sum_\sigma |s_{l\sigma}|^2 - 2S^{tot}_l \bigr|$.

\begin{figure}[tbp] 
\includegraphics[width=0.8\columnwidth,clip]{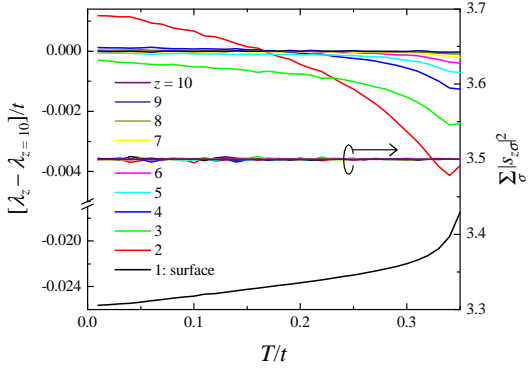} 
\caption{Layer dependent Lagrange multiplier $\lambda$ for the 20-layer system with a uniform transfer intensity, 
localized spin $S=3/2$, and average carrier density $n=0.5$. 
Kinks at $T \sim 0.34t$ indicate the spinor Bose condensation, i.e., the ferromagnetic transition. 
The right axis shows the mean boson density at each layer, showing that the numerical error originating from 
the conjugate gradient algorithm is small. } 
\label{fig:lambda} 
\end{figure}

\begin{figure}[tbp]
\includegraphics[width=1\columnwidth,clip]{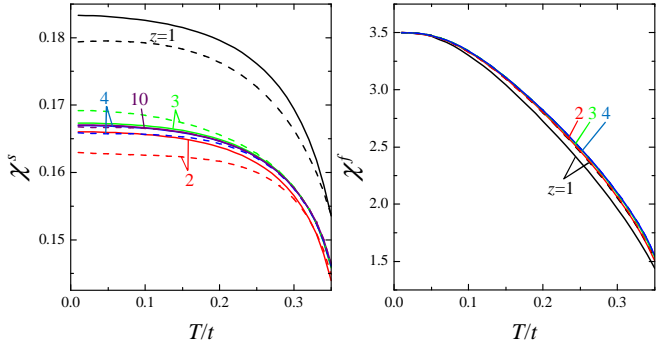} 
\caption{Order parameters $\chi_{ij}^s$ and $\chi_{ij}^f$ for the 20-layer DE model with uniform transfer. 
Solid lines are in-plane components $z=z_i=z_j$, while broken lines are out-of-plane components $z=z_i-1=z_j$. 
The order parameter $\chi^f$ converges to the bulk value more rapidly than $\chi^s$. } 
\label{fig:chi} 
\end{figure}

As an example, we show the results of $\lambda_i$ and $\sum_\sigma |s_{l\sigma}|^2$ in figure~\ref{fig:lambda} and 
order parameters $\chi^s$ and $\chi^f$ in figure~\ref{fig:chi}. 
The magnetization profile corresponding to this choice of parameters is shown in figure~\ref{fig:MofT}. 
The Lagrange multiplier $\lambda$ depends on the layer index $z$ and temperature $T$, while the boson density remains unchanged, 
indicating the accuracy of the conjugate gradient algorithm and applicability of the present Schwinger-boson method 
for spatially inhomogeneous systems. 
Typical error in the constraint was found to be less than 0.5~\% at each layer, and 
the error in $\chi^s$ and $\chi^f$ is much smaller. 

Another point to note is that our system is two dimensional. 
Therefore, strictly speaking, there is no Bose condensation (no magnetic ordering) at finite temperature 
unless there exists spin anisotropy. 
In this work, we discretize the momentum space and 
consider the lowest energy subband at $(k_x,k_y) =(0,0)$ with the wave function $\psi_0  = \sum_\sigma \sum_{l=1}^N a_l s_{l \sigma}$ 
as the Bose condensation.  
This corresponds to introducing a low-energy cutoff for spinor excitations representing coupling with the bulk region. 
In the following, we mainly take $\sqrt{2} \, 128 \times \sqrt{2} \, 128$ $k$ points in the first Brillouin zone, 
so the typical cutoff energy is $\chi^s t (\Delta k)^2 \approx 3 \times 10^{-5} t$. 
The magnetic transition temperatures of the bulk system and the layer system with $N=20$ were found to agree within $\sim 5~\%$. 
Thus, we believe that this method for a very thick system provides reasonable approximation of the surface behavior. 
The wave function $\psi_0$ determines the layer-dependent spinor condensation (the ordered magnetic moment). 

In layered systems, $n$ and $S^{tot}$ generally depend on the layer index. 
In this work, we consider an average carrier density $n = 0.5$. 
Particle-hole symmetry guarantees that $n_i$ and $S^{tot}_i$ are uniform at all temperatures. 
This choice does not lose generality as long as the system is well approximated by the DE model. 
The local spin is taken as $S=3/2$, therefore $S^{tot} = 1.75$

It is worth mentioning the shortcomings of the present SBMF method here. 
The bulk single-orbital DE model has been analyzed by using the DMFT 
and the ferromagnetic Curie temperature has been computed \cite{Furukawa95}. 
For a cubic lattice, the DMFT with the classical $t_{2g}$ spins predicts $T_C/t \sim 0.2$ at $x=0.5$, 
while the SBMF gives $T_C/t \sim 0.36$ \cite{OkamotoDE}.  
Thus, $T_C$ is about a factor 2 overestimated in the latter. 
This is probably because the carriers do not suffer from scattering due to the randomly distributed (fluctuating) spins in the SBMF. 
Note that the DMFT also tends to overestimate $T_C$ because it neglects spatial correlations. 
In the more realistic two-band DE model for manganites, the on-site Coulomb interaction becomes one of the sources to reduce $T_C$ \cite{Held00}. 
In light of these facts, we mainly focus on the surface magnetism relative to the bulk. 

\section{Results}

\begin{figure}[tbp] 
\includegraphics[width=1\columnwidth,clip]{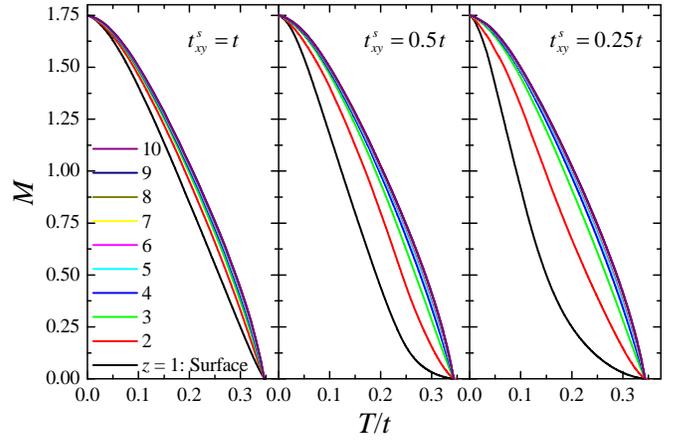} 
\caption{Layer dependent magnetization $M$ for the 20-layer DE model as a function of temperature $T$ using the Schwinger-boson mean-field approximation. 
Left panel: surface in-plane transfer is taken as $t^s_{xy}=t$, middle panel $t^s_{xy}=0.5t$, and right panel $t^s_{xy}=0.25t$. } 
\label{fig:MofT} 
\end{figure}

\begin{figure}[tbp] 
\includegraphics[width=0.9\columnwidth,clip]{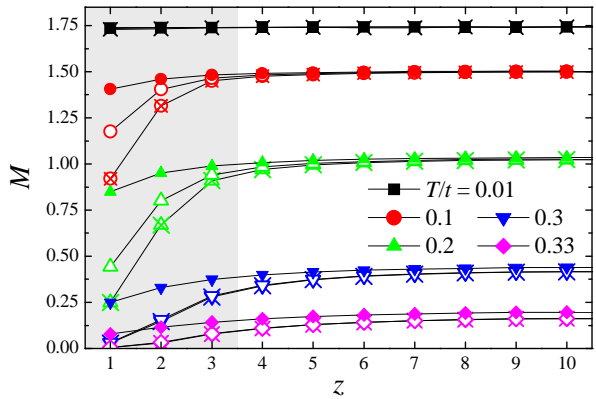} 
\caption{Layer dependent magnetization $M$ for the 20-layer DE model at various temperatures indicated. 
Filled (open,crossed) symbols are the results with the surface in-plane transfer $t^s_{xy}=t \,\, (0.5t, 0.25t)$. } 
\label{fig:MofZ} 
\end{figure}

Numerical results for the layer dependent magnetization of an $N=20$ layer system are shown in the left panel of figure~\ref{fig:MofT} 
as a function of temperature. 
The lattice constant is taken to be unity: surface layers are located at $z=1$ and 20, and 
the magnetization profile is symmetric with respect to the center of the system at $z=10.5$. 
Clearly, the surface magnetization decreases faster than the bulk magnetization with increasing temperature, 
but all magnetizations disappear at the same temperature. 
In figure~\ref{fig:MofT}, many lines are on top of each other, except for very small $z$. 
To see the layer dependence of the magnetization more clearly, 
we plot the magnetization as a function of the layer coordinate $z$ in figure~\ref{fig:MofZ} for various temperatures indicated. 
One can see that layers at $4 \le z \le 17$ show roughly the same magnetization, 
thus about three unit-cell wide surface layers is rather robust. 
Yet near $T_C$, the surface layer becomes thicker. 
Similar behavior is observed in the MC study for the Heisenberg model \cite{Binder74}. 

Next we discuss the effect of surface condition on the temperature dependent magnetization. 
Since we are using the simple single-orbital DE model, we simulate various effects by changing the transfer intensity around the surface layers. 
The ferromagnetic Curie temperature remains about $0.34t$ in all cases, 
indicating that $N=20$ is thick enough and the surface condition does not affect the bulk behavior. 

First, we reduce the surface intraplane transfer $t^s_{xy}$. 
This may correspond to either the surface roughness or the elongation of the MnO$_6$ octahedron stabilizing the $d_{3z^2-r^2}$ orbital in the surface layers. 
Results are shown in the middle and the right panels of figures~\ref{fig:MofT} and \ref{fig:MofZ}. 
In this case, the in-plane kinetic energy of the electrons is reduced in the surface layers. 
This causes rapid suppression of the magnetization. 
However, coupling between the surface layers and the bulk region induces finite magnetization on the surface, resulting in the long tail of the magnetization curve. 
Even in this case, layers at $4 \le z \le 17$ show roughly the same magnetization. 

\begin{figure}[tbp] 
\includegraphics[width=1\columnwidth,clip]{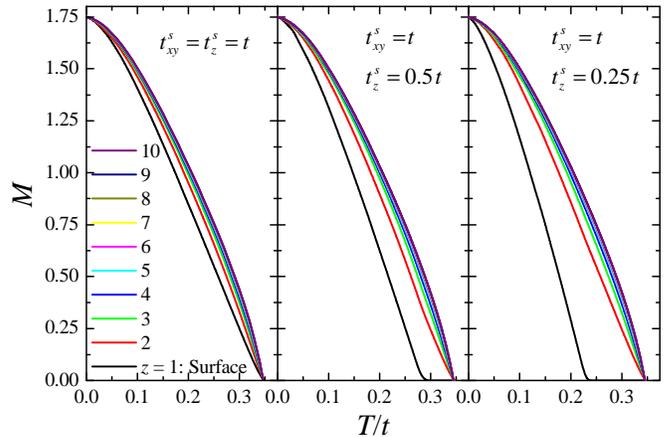} 
\caption{Layer dependent magnetization $M$ for the 20-layer DE model as a function of $T$ using the Schwinger-boson mean-field approximation. 
Left panel: uniform transfer $t$, middle panel $t^s_{z}=0.5t$, and right panel $t^s_{z}=0.25t$. } 
\label{fig:MofTtz} 
\end{figure}

\begin{figure}[tbp] 
\includegraphics[width=0.9\columnwidth,clip]{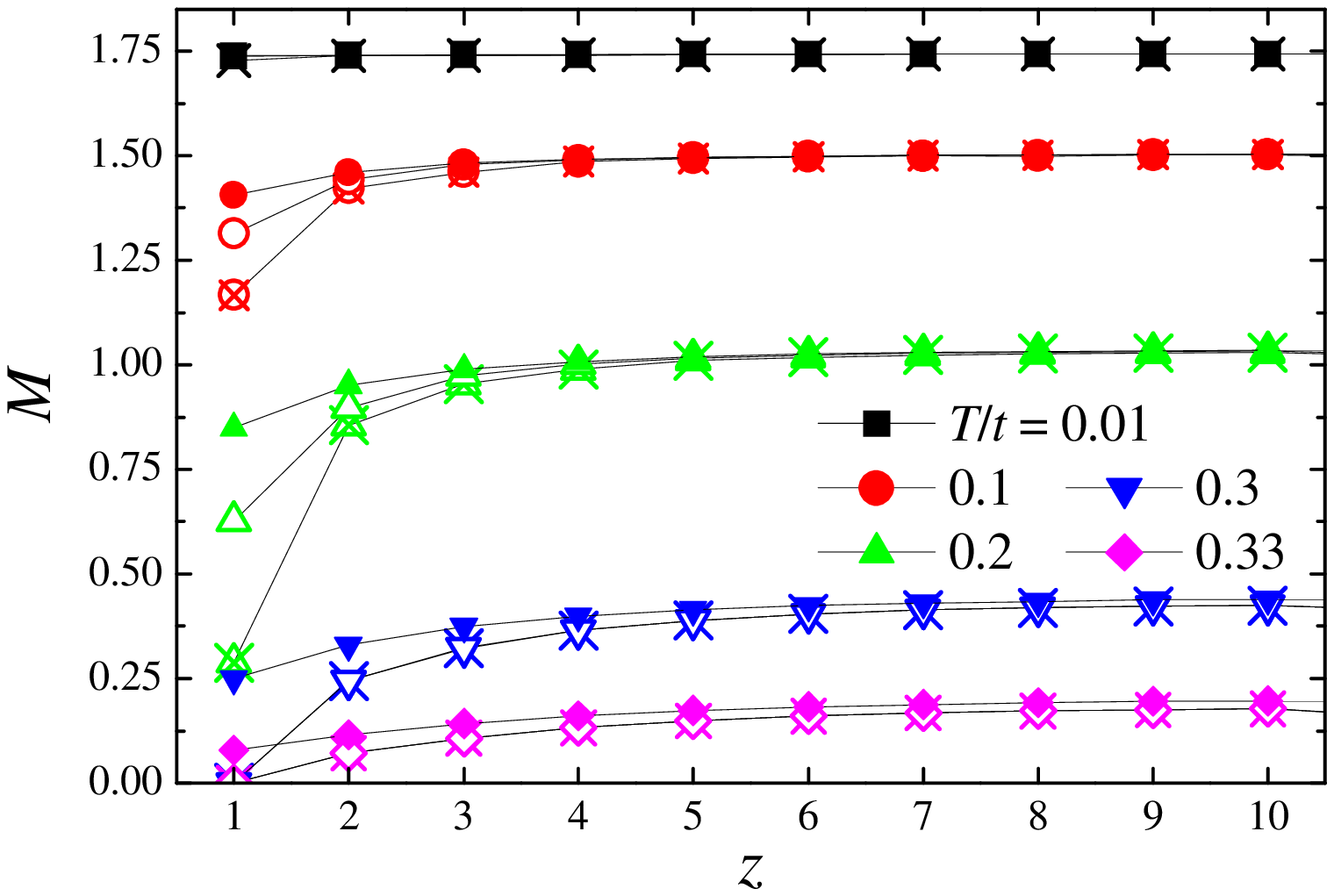} 
\caption{Layer dependent magnetization $M$ for the 20-layer DE model at various temperatures indicated. 
Filled (open,crossed) symbols are the results with the surface out-of-plane transfer $t^s_{z}=t \,\, (0.5t, 0.25t)$. } 
\label{fig:MofZtz} 
\end{figure}

\begin{figure}[tbp] 
\includegraphics[width=1\columnwidth,clip]{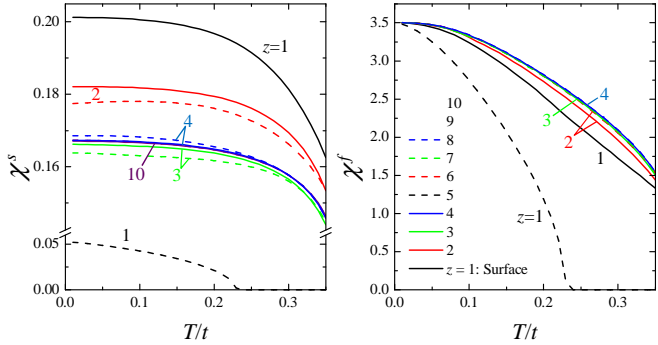} 
\caption{Order parameters $\chi_{ij}^s$ and $\chi_{ij}^f$ for the 20-layer DE model with the surface out-of-plane transfer $t_z^s = 0.25t$. 
Solid lines are in-plane components $z=z_i=z_j$, while broken lines are out-of-plane components $z=z_i-1=z_j$. } 
\label{fig:chitz025} 
\end{figure}

Second, the interlayer transfer $t^s_{z}$ between the surface layer and its neighboring layer is reduced. 
This roughly corresponds to the contraction of the MnO$_6$ octahedron in the surface layers, resulting in the increase of the $d_{x^2-y^2}$ orbital occupancy. 
Results are shown in the middle and the right panels of figures~\ref{fig:MofTtz} and \ref{fig:MofZtz}.
In contrast to the reduction of intraplane transfer, 
strong ferromagnetic correlations remain in the surface layers. 
This prevents the rapid reduction of the surface magnetization at low temperature. 
With increasing temperature, the interlayer magnetic correlation is rapidly reduced 
as can be seen in the temperature dependence of the order parameters $\chi^s$ and $\chi^f$ in figure~\ref{fig:chitz025}. 
Thus, the surface layer becomes more two dimensional. 
Eventually, surface magnetism disappears below the bulk $T_C$, accompanying the disappearance of $\chi^s$ and $\chi^f$ on 
the bonds connecting the surface layer and its neighboring layer. 
Above this transition, the rest of the system behaves like the one with `clean' surfaces located at $z=2$ and 19, since the surface layers are decoupled. 
The clear surface transition might be an artifact of the mean-field approximation, 
and in reality the tiny magnetization on the surface layers may survive. 
If the antiferromagnetic interaction between the local $t_{2g}$ spins (neglected in the present calculations) 
is strong relative to the interplane ferromagnetic correlation, 
the surface magnetic moment would change its relative orientation to the bulk moment.  
In either case, the small surface magnetic moment is expected to survive up to the true bulk $T_C$.

The surface magnetic moment of cubic manganites was reported in \cite{Park98}. 
It was shown that the surface moment is much smaller than that in the bulk, but disappears at the bulk $T_C$. 
The experimental result seems closest to the theoretical curve with $t^s_{xy} = 0.25 t$ in figure~\ref{fig:MofT}. 
This indicates that the interlayer ferromagnetic coupling remains near the surface. 
On the other hand in the bilayer manganites, the surface layer does not show a ferromagnetic moment, 
while the next-to-surface layer shows almost bulk-like magnetization \cite{Freeland05}. 
This situation may correspond to the weak interlayer coupling limit.  
In this case, the magnetization in the second layer is identical to that in the ideal surface. 
Therefore, the magnetization is closer to the bulk value, although it is reduced somewhat. 
But in the low $T_C$ systems, more complexity would exist due to various effects such as charge/orbital orderings absent in the present model.

\section{Discussion and summary}

Here, we would like to discuss how surface (interface) magnetization influences the TMR effect. 
We consider an ideal tunneling junction in which an interface between a ferromagnetic electrode and an insulating barrier is flat. 
Furthermore, the potential barrier is very high compared with the band width inside the electrodes. 
Thus, the interface layer of an electrode is equivalent to the surface layer in the previous discussion. 
The important quantity characterizing a TMR junction is the MR 
defined by ${\rm MR} = (G_P-G_{AP})/G_{AP}$ 
with $G_{P(AP)}$ the tunneling conductance with parallel (antiparallel) alignment of the magnetization of the electrodes \cite{MR}. 
When the dependence of the tunneling matrix on the Fermi velocity $\vec v$ is weak and 
two electrodes are identical, 
MR is expressed as 
\begin{equation}
{\rm MR} = \frac{2P^2}{1-P^2}, 
\label{eq:MR}
\end{equation}
where $P$ is the spin polarization at the Fermi level in an electrode defined by 
$P = (\rho_{\uparrow} - \rho_{\downarrow})/(\rho_{\uparrow} + \rho_{\downarrow})$ with 
$\rho_\sigma$ the density of states of electron with spin $\sigma$ at the Fermi level. 
Note, MR diverges when $P=1$ (full polarization). 
Therefore, it is not surprising to have very large MR for manganites at low temperature, as observed experimentally \cite{Bowen03}. 
In general, the assumptions to arrive at equation~(\ref{eq:MR}) are not satisfied, 
and one has to consider the realistic band structure in the presence of the interfaces 
and the dependence of the tunneling current on the barrier height and thickness. 
However, for a simple free electron model, it was shown that the exact solution of MR approaches the Julli{\`e}re's model 
as barrier thickness and height increase \cite{MacLaren97}. 
Having these facts in mind, following discussion will be done on a qualitative level.

\begin{figure}[tbp] 
\includegraphics[width=0.8\columnwidth,clip]{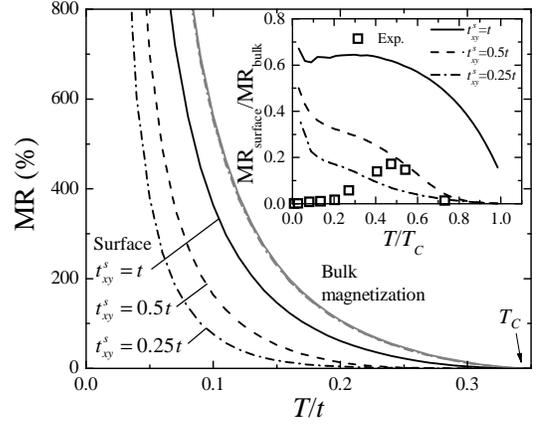} 
\caption{Temperature dependence of MR computed
by equation~(\ref{eq:MR}) using surface magnetization (black lines, MR$_{\rm surface}$) and bulk magnetization (gray lines, MR$_{\rm bulk}$). 
Solid (dashed, dash-dotted) lines are the results with the surface in-plane transfer $t_{xy}^s = t \,\, (0.5t,0.25t)$. 
Inset: the ratio between MR$_{\rm surface}$ and MR$_{\rm bulk}$. 
Experimental MR$_{\rm surface}$ are taken from \cite{Bowen03}, and 
experimental MR$_{\rm bulk}$ is estimated using equation~(\ref{eq:MR}) with $M$ from \cite{Urushibara95} for $x=0.3$. }
\label{fig:MR} 
\end{figure}

Figure \ref{fig:MR} summarizes the results for MR using equation~(\ref{eq:MR}). 
Instead of the spin polarization at the Fermi level, we use the total polarization $P = M/S^{tot}$. 
For the bulk double-exchange model, this is a rather good approximation. 
It is clearly shown that a smaller polarization at a surface layer reduces MR substantially. 
This tendency becomes stronger when the surface transfer is reduced. 
Experimental results reported in \cite{Bowen03} shows that the MR becomes as large as 800~\% at the lowest temperature, but disappears far below $T_C$. 
This can be understood by the interface magnetization, which dominates the MR properties. 
Since we cannot make a direct comparison between the absolute values of the theoretical MR and the experimental one, 
let us consider the ratio between the actual MR and the MR expected from the bulk magnetization 
denoted by MR$_{\rm surface}$ and MR$_{\rm bulk}$, respectively. 
To compute the experimental ratio, the MR data in figure 3(b), \cite{Bowen03}, is used for MR$_{\rm surface}$, 
and MR$_{\rm bulk}$ is computed using equation~(\ref{eq:MR}) with $M$ taken from \cite{Urushibara95} for the carrier concentration $x=0.3$. 
Here, $M$ is assumed to be fully saturated at the lowest temperature. 
The result is presented in the inset of figure~\ref{fig:MR}. 
It is shown that the experimental ratio MR$_{\rm surface}/$MR$_{\rm bulk}$ is comparable to the theoretical one for $t_{xy}^s=0.5t$ at temperatures above $\sim 0.5 T_C$. 
This may suggest that the quality of the experimental interfaces is rather high and the complexity inherent in doped manganites, 
such as charge-orbital ordering and (chemical) phase separation, is not so important near $T_C$. 
Note that in \cite{Bowen03} nearly optimally-doped manganites ($x=1/3$) are used. 
With decreasing temperature, the theoretical MR increases and diverges at $T=0$, and therefore MR$_{\rm surface}/$MR$_{\rm bulk}$ approaches 1. 
On the other hand, the experimental MR saturates at low temperature and, therefore, MR$_{\rm surface}/$MR$_{\rm bulk}$ decreases. 
This indicates that, in the experiment, additional interactions, such as the antiferromagnetic superexchange interaction between the localized $t_{2g}$ spins, 
compete with the DE interaction at low temperatures. 
The resulting canted spin structure would reduce the ferromagnetic polarization and suppress the divergence of MR.

\begin{figure}[tbp] 
\includegraphics[width=0.9\columnwidth,clip]{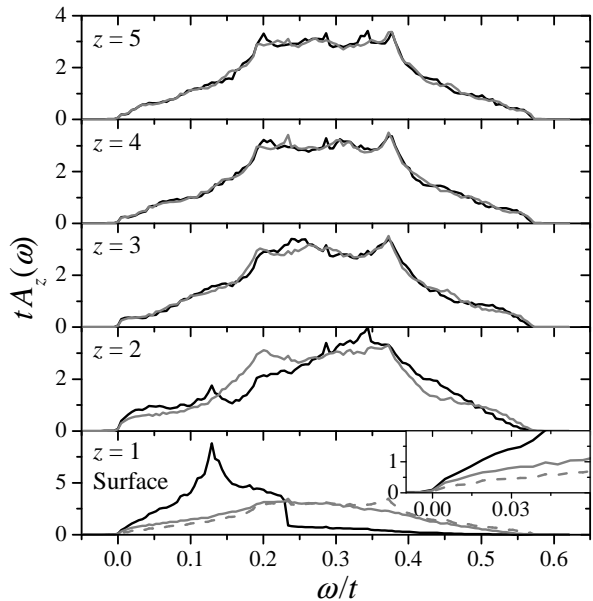} 
\caption{Layer dependent spectral function of spinor bosons computed at $T=0.01t$. 
The small imaginary number $i \eta$ with $\eta = 0.01 \chi^s / 2 S^{tot} \approx 4 \times 10^{-4} t$ is introduced in the lattice Green's function. 
Black and gray lines are the results with the surface in-plane transfer $t^s_{xy}=0.5t$ and $t$, respectively. 
For comparison, the spectral function at $z =10$ with $t^s_{xy}=t$ is also shown as a gray dashed line in the lowest panel. 
Inset: a magnified view in the low energy region. }
\label{fig:spectra} 
\end{figure}

Faster suppression of the surface magnetization in the theoretical results is ascribed to the larger thermal fluctuation of spins. 
To see this effect more clearly, 
we computed the layer dependent spectral function $A_z (\omega)$ of spinor bosons. 
$A_z (\omega)$ is defined as 
$A_z (\omega) = - \frac{1}{\pi} \Im \!\int \!\! \frac{dk^2}{(2 \pi)^2} \{ \omega + i \eta - H_s (k_x,k_y) \}^{-1}|_{zz}$ with 
$\omega$ a real frequency and $i \eta$ a small imaginary number. 
$H_s$ is the mean-field Hamiltonian for the spinor bosons given by the second and the third terms in $L_s$. 
$A_z (\omega)$ for $s_\uparrow$ and $s_\downarrow$ have the same spectral shape. 
Numerical results for $N=20$ systems are presented in figure~\ref{fig:spectra}. 
In the bulk region, $\chi^s \sim 0.167$, so the full band width of the boson excitation is $12 \chi^s t /2S^{tot} \approx 0.573 t$. 
As can be seen, low-energy fluctuations are largely enhanced at surface layers $z=1$ compared with those in the bulk. 
Even in the uniform transfer case (surface $t^s_{xy} = t$), 
the low-energy part of the surface spectral function is about twice as large as that of the bulk ($z=10$). 
Therefore, thermal excitation of spinor bosons has a stronger effect, resulting in 
the rapid suppression of the magnetization. 
It is also shown that the spectral function starts to develop the bulk-like shape at $z \sim 4$. 
This explains why surface layers are about three-unit-cell wide.

The surface spectral function shown in figure~\ref{fig:spectra} may suggest a possible way to keep a large polarization at high temperature 
to improve MR; 
thereby suppressing the low energy spin fluctuations.  
This may be achieved by, for example, 
(1) creating uniaxial spin anisotropy or 
(2) using a magnetic insulator with a relatively high ordering temperature as a barrier. 
For (1), applying a slightly compressive strain for manganites would work. 
For (2), a possible candidate is BiFeO$_3$ (with [111] stacking). 
In this case, an exchange-bias-type effect is also expected. 
A parent compound of the manganites, LaMnO$_3$, with [001] stacking may not be helpful because of its low N{\'e}el temperature.

The present model includes only DE interactions. 
Therefore, at the lowest temperature, all spinor bosons are condensed at $(k_x,k_y) = (0,0)$ and the ferromagnetic moment is saturated. 
In reality, additional interactions may create complexity. 
For example, surface polarity and segregation \cite{Choi99,Bertacco02} are expected to change the surface carrier density and reduce the ferromagnetic interaction. 
Therefore, antiferromagnetic superexchange interactions between the localized $t_{2g}$ spins 
are expected to reduce the ferromagnetic polarization at low temperature. 
Further complexity, such as polaronic effects and charge-orbital ordering, would enhance the effect of the superexchange interactions. 
The present SBMF method is rather simple, and including such complexities to the fermionic part is possible. 
We are currently working on including the electron-lattice couplings with orbital degeneracy using a dynamical-mean-field-type treatment 
to study the magnetic and metal-insulator transitions at the surface and in thin films \cite{Hong05}. 

To summarize, we studied the surface magnetic behavior of the double-exchange model for doped manganites using the Schwinger-boson mean-field method. 
Low-energy spin fluctuations are enhanced at the surface and the magnetic moment is suppressed more rapidly than in the bulk. 
We further considered an ideal tunneling magnetoresistance junction consisting of two manganite leads and an insulating barrier. 
The magnetoresistance ratio of such a junction is determined by the surface polarization. 
Therefore, it decreases much faster than the bulk magnetization when the temperature is increased. 
A possible improvement of the MR is expected from suppressing the low-energy spin fluctuations.


\section*{Acknowledgments}
The author acknowledges useful discussions with K. Bevan, R. Fishman and Z. Y. Zhang. 
This work was supported by the Division of Materials Sciences and Engineering, Office of Basic Energy Sciences, U.S. Department of Energy.



\begin{thebibliography}{99}
\bibitem{MR}There are different definitions for MR depending on how to normalize the relative conductance.
\end{thebibliography}
\end{document}